\documentclass[10pt,conference,letterpaper]{IEEEtran}

\usepackage{cite}
\usepackage{amsmath,amssymb,amsfonts}
\usepackage{algorithmic}
\usepackage{graphicx}
\usepackage{textcomp}
\usepackage{xcolor}
\usepackage[hidelinks]{hyperref}
\usepackage{booktabs}
\usepackage{multirow}
\usepackage{fancyhdr}
\usepackage{tikz}

\makeatletter
\def\ps@IEEEtitlepagestyle{%
  \def\@oddfoot{\hspace*{-0.6cm}\footnotesize
  979-8-3315-9314\hspace{0.3em}8/25/\$31.00~\copyright~2025 IEEE}%
  \def\@evenfoot{}%
}
\makeatother
\title{KNN and Time Series Based Prediction of Power Generation from Renewable Resources}
\author{
    Ismum Ul Hossain\\
    Department of Electrical and Computer Engineering\\
    University of Hawaii at Manoa, USA\\
    siamjosephite43@gmail.com
    \and
    Mohammad Nahidul Islam \\
    Department of Chemical Engineering\\
    KUET, Bangladesh\\
    nahidulislam.che@gmail.com
}

\begin{document}

% ----- Override IEEE title page footer -----
\makeatletter
\def\ps@IEEEtitlepagestyle{%
  \def\@oddfoot{\footnotesize 979-8-3315-9314 8/25/\$31.00 ©2025 IEEE\hspace{1cm}}%
  \def\@evenfoot{}%
}
\makeatother
% ------------------------------------------

\maketitle

% -------- HEADER FOR FIRST PAGE --------
\thispagestyle{fancy}
\fancyhf{}
\fancyhead[L]{\footnotesize
2025 IEEE International Conference on Power, Electrical, Electronics and Industrial Applications (PEEIACON)\\
04--05 December 2025, Southeast University (SEU), Dhaka, Bangladesh}
\renewcommand{\headrulewidth}{0pt}

\begin{abstract}

As the world shifts towards utilizing natural resources for electricity generation, there is need to enhance forecasting systems to guarantee a stable electricity provision and to incorporate the generated power into the network systems. This work provides a machine learning environment for renewable energy forecasting that prevents the flaws which are usually experienced in the actual process; intermittency, non-linearity and intricacy in nature which is difficult to grasp by ordinary existing forecasting procedures. Leveraging a comprehensive approximately 30-year dataset encompassing multiple renewable energy sources, our research evaluates two distinct approaches: K-Nearest Neighbors (KNN) model and Non-Linear Autoregressive distributed called with Seasonal Autoregressive Integrated Moving Average (SARIMA) model to forecast total power generation using the solar, wind, and hydroelectric resources. The framework uses high temporal resolution and multiple parameters of the environment to improve the predictions. The fact that both the models in terms of error metrics were equally significant and had some unique tendencies at certain circumstances. The long history allows for better model calibration of temporal fluctuations and seasonal and climatic effects on power generation. The reliability enhancement in the prediction function, which benefits from 30 years of data, has value to grid operators, energy traders, and those establishing renewable energy policies and standards concerning reliability.
\end{abstract}

\begin{IEEEkeywords}
Renewable Energy, Forecasting, Machine Learning, KNN regression
\end{IEEEkeywords}

\section{Introduction}
Currently, fossil fuels constitute the predominant global energy source. Fossil fuels, comprising coal, oil, and natural gas, are hydrocarbons or their derivatives. Fossil fuels develop over millions of years; however, their reserves are diminishing more rapidly than they are being generated. Moreover, fossil fuels release greenhouse gases, increasing climate change and threatening the environment upon which humanity depends. The recent demand for clean and sustainable energy has intensified the emphasis on renewable energy research and development \cite{ref1}. In recent years, research and development in renewable energy have gathered considerable interest due to an increasing demand for clean and sustainable energy \cite{ref2}. Furthermore, renewable energy sources have numerous benefits, such as less reliance on foreign nations, employment generation, and the possibility of cost reductions \cite{ref3,ref4}. In addition, renewable energy can remove the utilization of fossil fuels and gain the vision of maintaining the environmental ecology. The intrinsic variability and uncertainty of renewable energy sources present considerable obstacles to their extensive adoption and integration into power systems \cite{ref5}. One way to address the issues linked with the nature of energy is by creating precise prediction models for energy productions. These models are essential for lessening the effects of energy supply on the power grid and aiding in improved planning and resource management \cite{ref6}.

According to the literature, China is facing the rising simultaneous difficulties of energy supply and demand, with rapid economic growth and substantial energy usage \cite{ref7} \cite{ref8} \cite{ref9}. The second-biggest economy in the world and most populous country, China, has a difficult time meeting its energy needs while maintaining environmental sustainability. Nonetheless, the nation has become a global pioneer in the production of renewable energy, offering an intriguing case study in the shift to clean energy. To provide a complete view, this section examines China's current renewable energy environment, examining its leading sources, legislative frameworks, and potential futures with pertinent data. China relies heavily on coal for electricity generation, but renewables are gaining ground. Coal accounted for 62\% of electricity in 2021, while renewable sources like wind and solar saw a significant rise in 2023 \cite{ref10}. Despite a massive installed coal capacity, China boasts a growing renewable capacity, reaching 1.26 TW in 2023, with wind and solar leading the charge \cite{ref11}. Nuclear power is also playing an increasing role, with over 5\% of electricity coming from nuclear sources and significant expansion plans in place \cite{ref12}.

China leads the world in both photovoltaic (PV) and solar thermal markets. After beginning domestic PV manufacturing in the late 1990s, strong government incentives in 2011 accelerated expansion, making China the top PV installer by 2013 and the largest PV producer by 2015 \cite{ref8}. China surpassed 100\,GW of PV capacity in 2017 and reached 253\,GW by 2020—one-third of global installations. Most PV output comes from western regions, supported by mega-projects such as the Huanghe Hydropower Golmud Solar Park and the Tengger Desert Solar Park \cite{ref12}. The Huanghe Hydropower Hainan Solar Park, with 2.2\,GW capacity, is the world’s second-largest \cite{ref13}. In 2023, China built the world’s largest hydro-solar plant in Sichuan. Despite its scale, solar provided only 3.5\% of national energy capacity in 2020, though China targets 1,200\,GW of combined solar and wind by 2030. Solar thermal remains significant, with 290\,GW$_{th}$ installed by 2014. By 2023, solar became cheaper than coal, and China added 45.7\,GW of PV in early 2024, a 34\% increase despite slower growth \cite{ref14}.

China has extensive wind resources, making wind energy its third-largest electricity source by 2021, contributing 7.5\% of national power generation. The country added 71.6 \,GW of wind capacity in 2020, reaching 281 \,GW—far ahead of the United States (118 \,GW). In 2021, China accounted for nearly 70\% of global new installations, followed by the U.S.\ (14\%) and Brazil (7\%) \cite{ref11}. By 2022, China became the world’s leading hub for wind-power equipment manufacturing. Goldwind, founded in 1998, expanded rapidly before its market share declined from 35\% in 2006 to 19\% in 2012. In 2019, Goldwind joined a wind-to-hydrogen project in northeast China to utilize excess wind energy \cite{ref12}.

\begin{figure}[!t]
\centering
\includegraphics[width=0.48\textwidth]{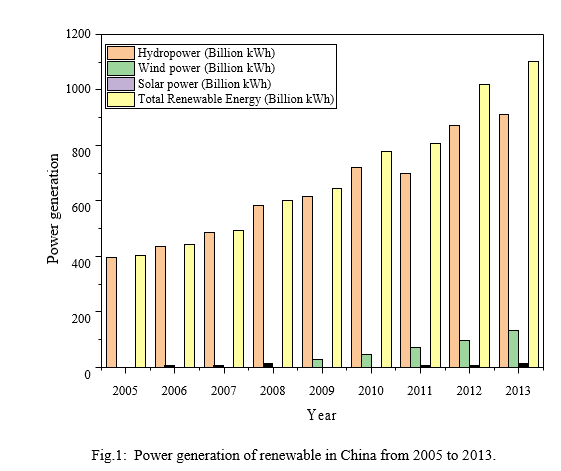}
\caption{Power generation of renewable in China from 2005 to 2013.}
\label{fig1}
\end{figure}

China Longyuan Electric Power Group Corp., a subsidiary of China Guodian Corporation, was an early leader in wind farm operations, managing 40\% of China's wind farms at one point \cite{ref15}. Hydroelectricity is China's largest source of renewable energy and second only to coal in overall energy production. As of 2021, China is the world's leading hydroelectricity producer, with an installed capacity of 390.9 GW, including 36.4 GW of pumped storage \cite{ref11}. This is a significant increase from 233 GW in 2011. In 2021, hydropower in China generated 1,300 TWh, up from 1,232 TWh in 2018, contributing about 18\% of the nation's total electricity generation \cite{ref16}.

\begin{figure*}[!t]
\centering
\includegraphics[width=0.65\textwidth]{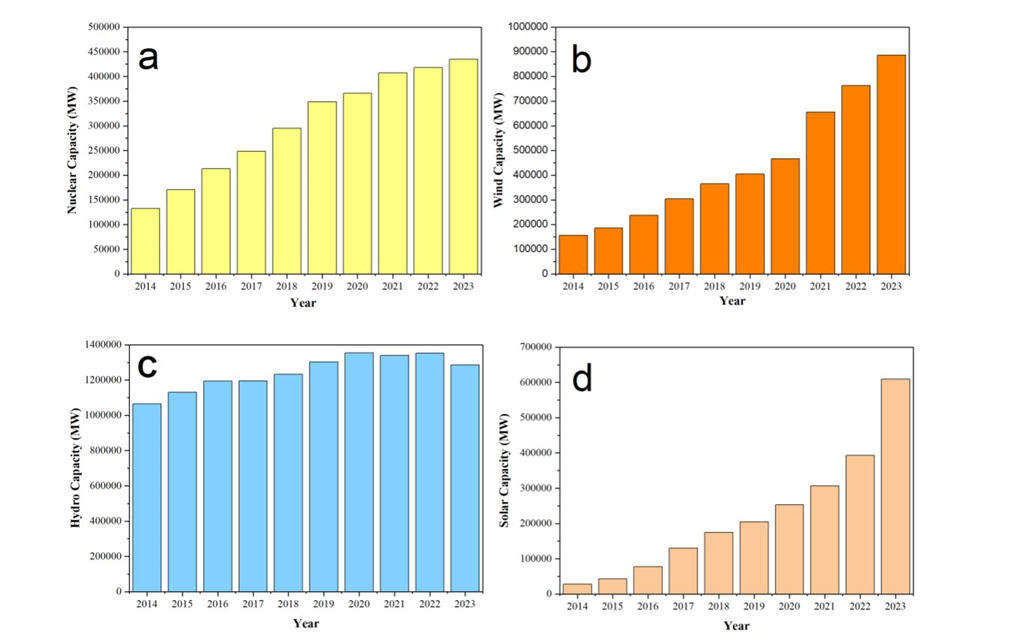}
\caption{Plot for a) Nuclear capacity of China from 2014 to 2023; b) Wind capacity of China from 2014 to 2023 \cite{ref14}; c) Hydro capacity of China from 2014 to 2023 \cite{ref20}; d) Solar capacity of China from 2014 to 2023 \cite{ref13}.}
\label{fig2}
\end{figure*}

China is a leading producer of nuclear power, ranking third globally in both installed capacity and electricity generation from nuclear energy, which accounts for about 10\% of the world's nuclear power output \cite{ref15}. As of February 2023, China operates 55 nuclear plants with a combined capacity of 57 GW, has 22 plants under construction adding another 24 GW, and plans over 70 more with a projected capacity of 88 GW \cite{ref17}. Nuclear energy provides around 5\% of China's electricity, generating 417 TWh in 2022. This is up from 53 reactors and 55.6 GW capacity in September 2022 \cite{ref18}. In 2019, nuclear power contributed 4.9\% to China's total electricity production with 348.1 TWh. Due to concerns about air quality, climate change, and fossil fuel shortages, China is considering nuclear power as a key alternative to coal. The China General Nuclear Power Group aims to reach 200 GW by 2035 with 150 new reactors \cite{ref19}. The objective of this paper is to introduce the power generation prediction for China with multiple techniques including K-Nearest Neighbors (KNN), Seasonal Autoregressive Integrated Moving Average (SARIMA). KNN, SARIMA are measured, providing superior results in terms of different error matrix.

\section{Methodology}
\begin{figure}[!t]
\centering
\includegraphics[width=0.45\textwidth]{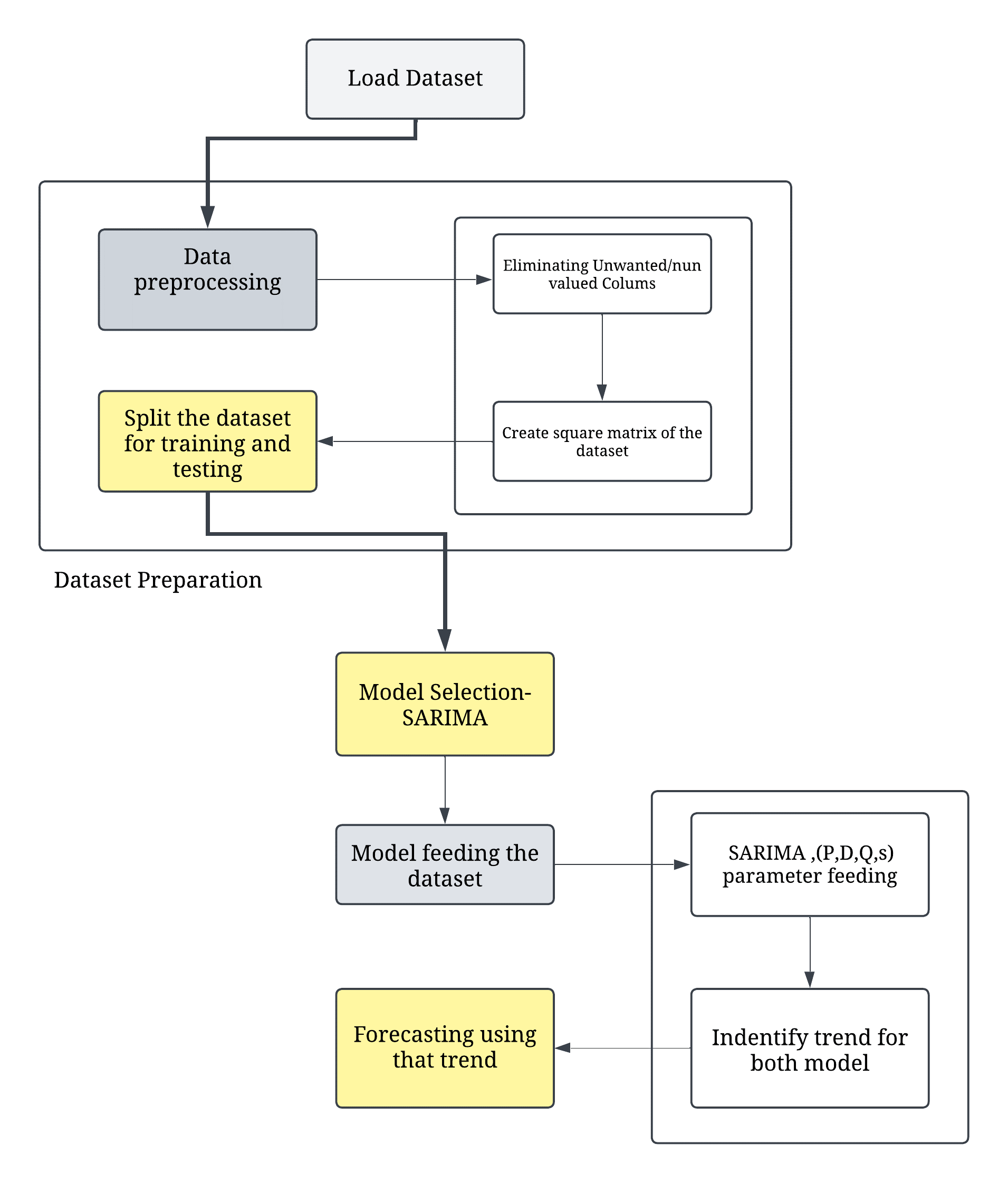}
\caption{Flowchart for model development and solution procedure.}
\label{fig3}
\end{figure}

In this article, the from data collection to prediction using time series algorithm have several steps. Each step is associated with each other with some sub steps in it. Fig. \ref{fig3} states the procedure of this research.

\subsection{Dataset Preparation}
Dataset collection was one of the major challenges of this research. Data was obtained from free sources such as the World Bank and IEA. The Excel file contained many power-generation columns, some of which had missing or inconsistent values; these were removed to create a clean square matrix where rows represent years and columns represent power generation from sources such as coal, gas, and renewables. The dataset was then suitable for model training, with 80\% used for training and the remaining 20\% for testing. For example, for data from 1968--2023, the training set covers 1968--2012, and the rest is used for testing.

\subsection{Feeding the Model}

\begin{figure}[!t]
\centering
\includegraphics[width=0.45\textwidth]{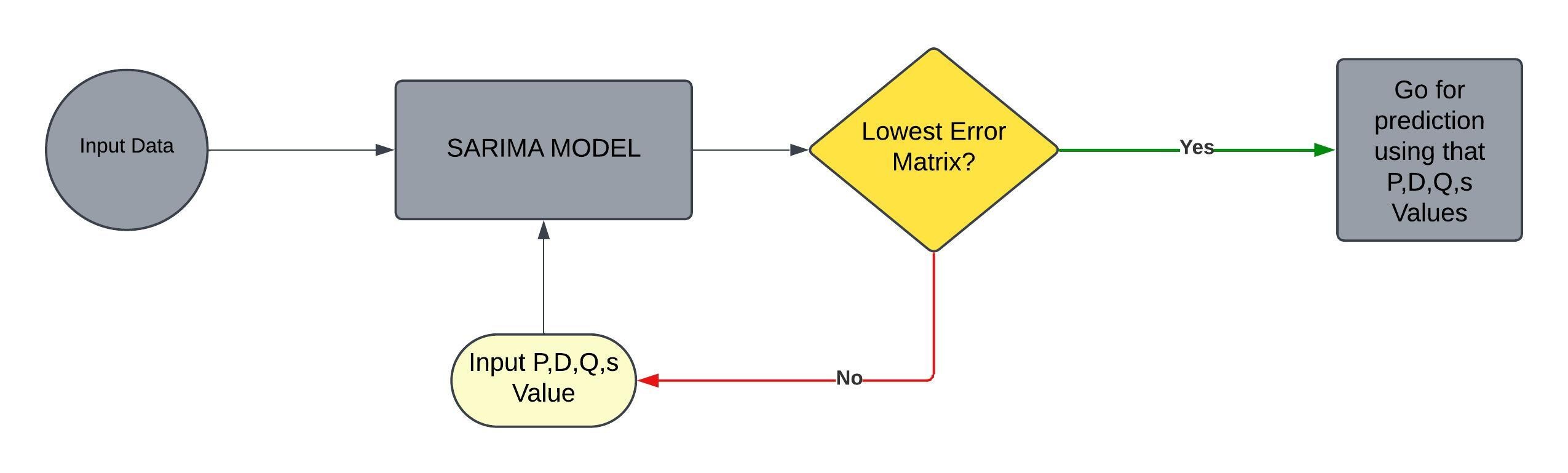}
\caption{Visualization of SARIMA Model Loop.}
\label{fig4}
\end{figure}

\begin{figure}[!t]
\centering
\includegraphics[width=0.48\textwidth]{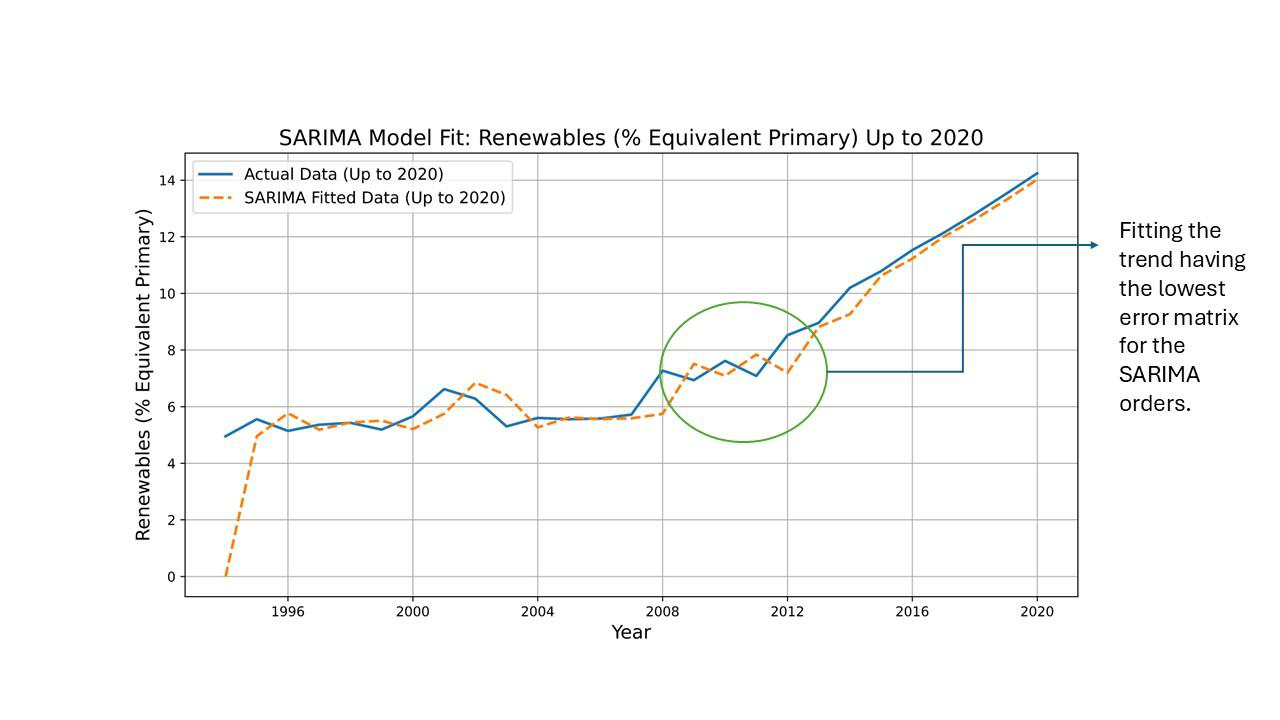}
\caption{Actual data vs. SARIMA-fitted values up to 2020, highlighting model precision in capturing trends with minimal residual error.}
\label{fig5}
\end{figure}

ARIMA and SARIMA are time series models with parameters P, D, Q, and s that must be selected to train the dataset. For each combination of P, D, Q (ARIMA) and P, D, Q, s (SARIMA), the training and output graphs would differ. Visualizing the output graph can reveal which curve matches the existing graph, lowering MSE and MAE. The dataset was split into 80\% training and 20\% testing to validate the model. After achieving the lowest MSE MAE, the remaining predictions were made using ARIMA and SARIMA parameters P, D, Q, s. Fig. \ref{fig5} displays the ARIMA Model parametric value selection diagram. ARIMA-like for SARIMA. The SARIMA model decomposed the historical share of renewable energy (\% equivalent primary) and projected its future expansion. Because it can model seasonal and trend data on renewable energy usage, SARIMA was utilised. Fig. \ref{fig5} shows how the SARIMA model captured data through 2020. Actual data, plotted in blue, came from renewable energy share datasets. The orange dashed line represents SARIMA-fitted data points to determine whether the model captures dataset patterns. The grey shaded area is based on these. The best method was used to determine SARIMA parameters for systematic estimation. Different models of seasonal and non-seasonal integrated orders (p, d, q) $\times$ (P, D, Q, s) were compared based on performance criteria like AIC and BIC values. The optimum configuration was obtained by choosing the residual error least different from the other set while achieving high accuracy and representation requirements. The model reflects the evolving periodicity of renewable energy adoption over a transitional period from 2008 to 2012, as shown by the green circle on the graph. This illustrates that the SARIMA model can account for daily changes and case growth. This fitted model will help predict future renewable energy shares, which will guide renewable energy development scenarios and strategic planning. However, separating the data into training and test data sets to validate the model will be its greatest strength in predicting the next array items.

The diagram below Fig. \ref{fig6} shows the applied methodological map of the case study with the focus on K-Nearest Neighbors (KNN) as the algorithm for prediction and data analysis. The stepwise process is described below:

\begin{figure}[!t]
\centering
\includegraphics[width=0.20\textwidth]{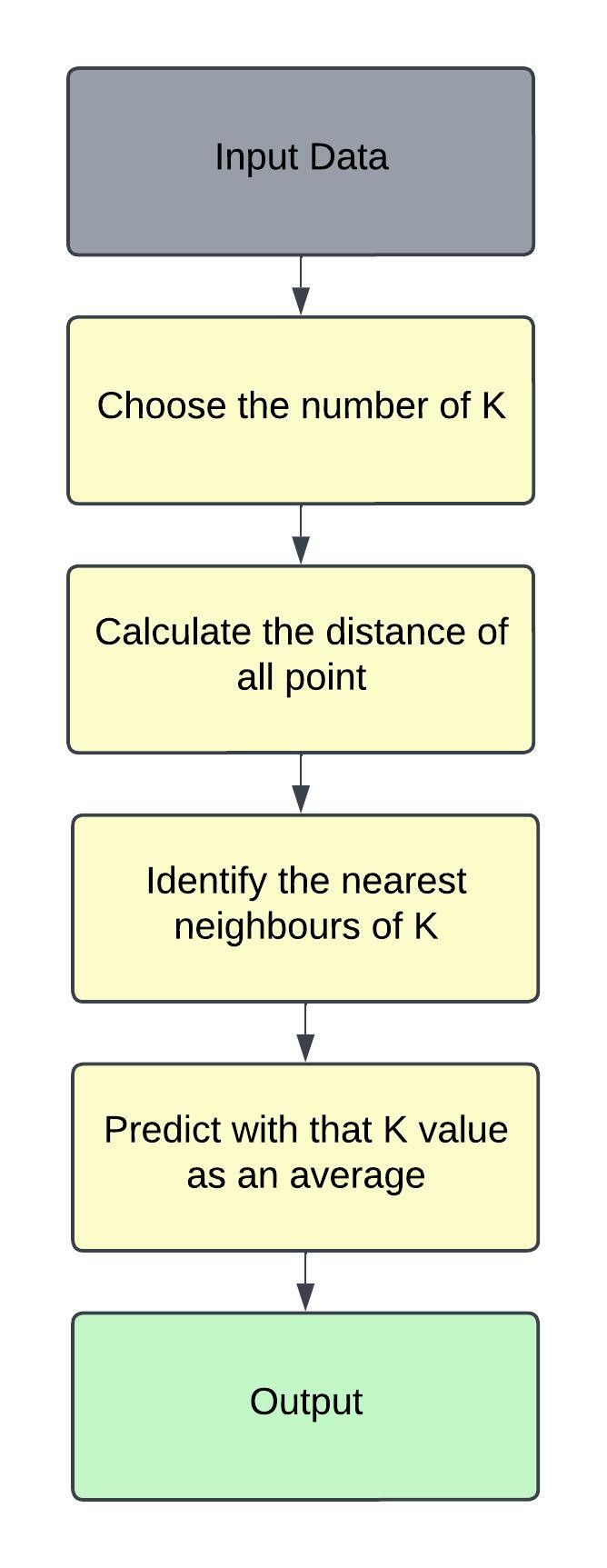}
\caption{Block diagram of KNN methodology.}
\label{fig6}
\end{figure}

\subsection{Forecasting the Data}
A forecast for the subsequent ten years was then made from the historical trend employing the optimized SARIMA model. After getting the optimized parameters in SARIMA model (P, D, Q) and (s) for better MSE and MAE to achieve better prediction the further prediction has been made. This model proved useful in quantifying cyclical and trended factors that produced reasonable forecasts based on prior data. Further, the K-Nearest Neighbors (KNN) algorithm was used as a second strategy to predict future values of the parameter of interest. As for KNN, an evaluation of the number of neighbors was done with cross validation and prediction is done by averaging the (K) nearest data points. The projection of KNN was done and compared with the projections of SARIMA for the validation of these trends and values. In both approaches, these methodologies provided a broad coverage of the expected trajectory in the next 10 years, taking advantage of the temporal analysis while both having a lower error matrix.

\section{Result and Analysis}

\subsection{KNN Prediction Scenario}
This section goes further and present the trend and forecast of the renewable energy by using function generated by the KNN for Hydro, Solar, Wind and of course, Total Renewables, both in electricity and equivalent primary energy. These graphs show the historical period data for the years 1992--2024 and a projection for the following years, 2025--2031.

\begin{figure}[!t]
\centering
\includegraphics[width=0.48\textwidth]{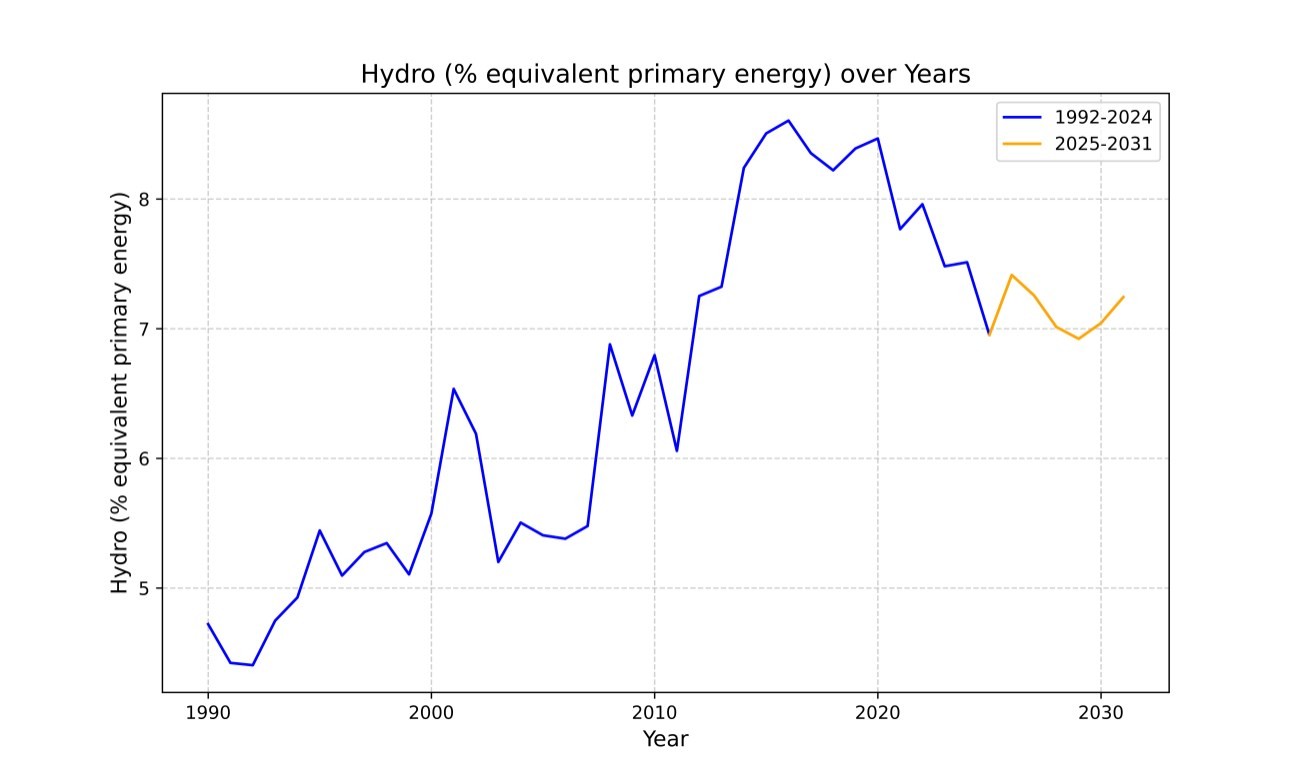}
\caption{Hydro (\% Equivalent primary energy) Prediction Scenario.}
\label{fig8}
\end{figure}

\textit{Hydro (\% Equivalent primary energy):} The contribution of hydro to primary energy as depicted in Fig. \ref{fig8} is stable throughout between 1992--2024 a slight initial steep drop between 2025 and 2031. The change highlighted its main form of renewable energy in hydro resource which may in future gradually shift to other renewable energy sources.

\begin{figure}[!t]
\centering
\includegraphics[width=0.48\textwidth]{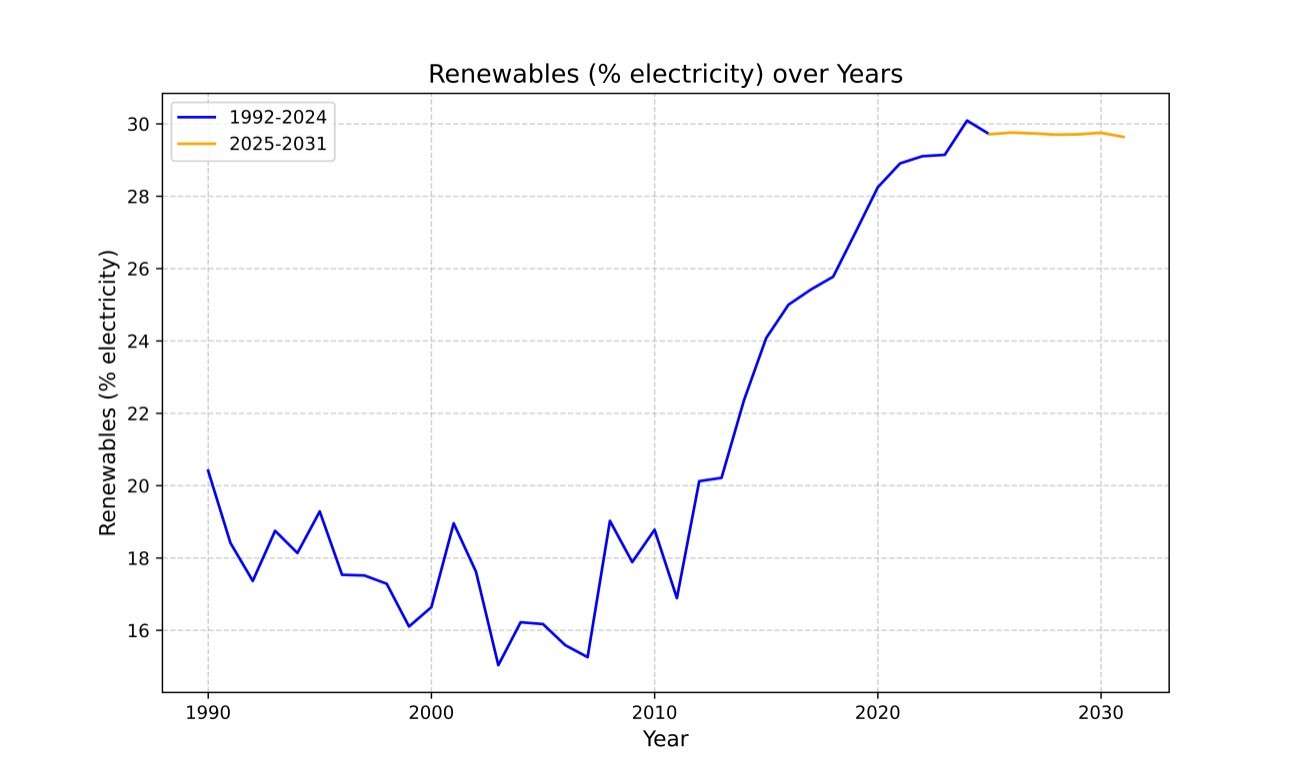}
\caption{Renewables (\% Electricity) Prediction Scenario.}
\label{fig9}
\end{figure}

\textit{Renewables (\% Electricity):} The Fig. \ref{fig9} shows an upward trend in the proportion of renewable energy source used in electricity generation more especially after the year 2010. After 2025, they believe the figures will remain relatively stagnant at approximately 30\% annually. This stabilization coincides with the global switch toward renewable resources, suggesting the development of a sturdy frame for this kind of infrastructure.

\begin{figure}[!t]
\centering
\includegraphics[width=0.48\textwidth]{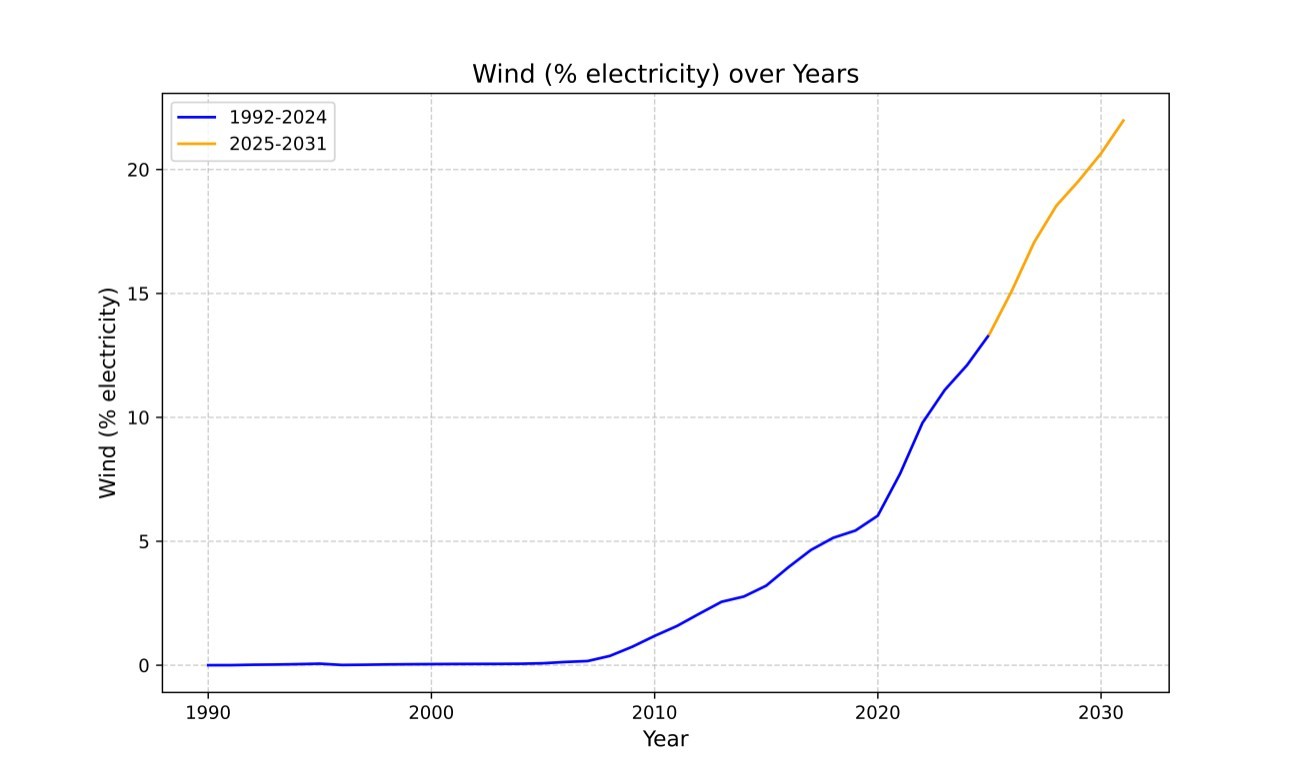}
\caption{Wind (\% equivalent primary energy) Prediction Scenario.}
\label{fig10}
\end{figure}

\textit{Wind (\% Electricity):} The wind power share of generation has increased gradually from 2005 to a projected level of more than 20 per cent in 2031 based on the forecast beyond 2025 as shown in Fig. \ref{fig10}. This growth demonstrates the capacity for expansion of wind technology and the appropriateness of its geographic setting.

\begin{figure}[!t]
\centering
\includegraphics[width=0.48\textwidth]{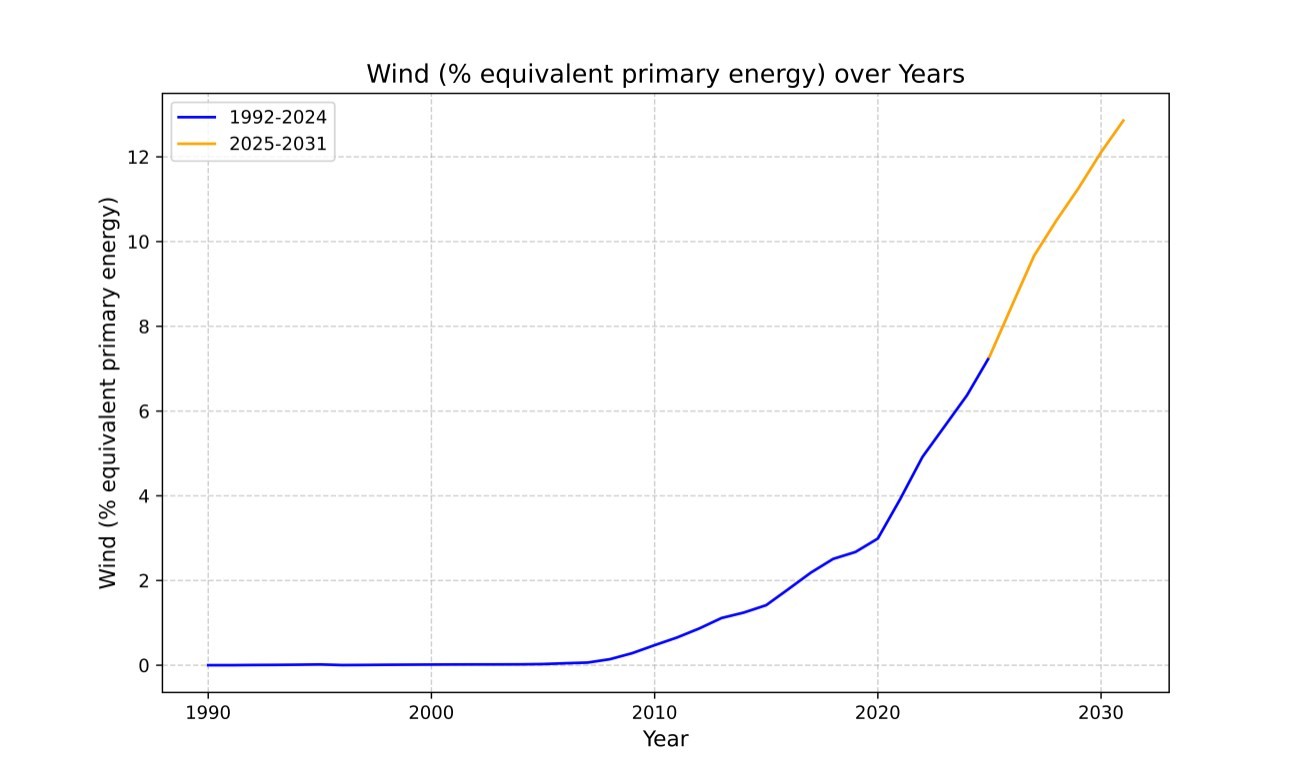}
\caption{Wind (\% equivalent primary energy)Prediction Scenario.}
\label{fig11}
\end{figure}

\textit{Wind ((\%) equivalent to primary energy):} A similar trend is shown in Fig. \ref{fig11} by the equivalent primary energy graph in which wind power is expected to be contributing over 12\% by 2031. It also affirms its predicate to fill the extra capacity from other renewable resources.

\begin{figure}[!t]
\centering
\includegraphics[width=0.48\textwidth]{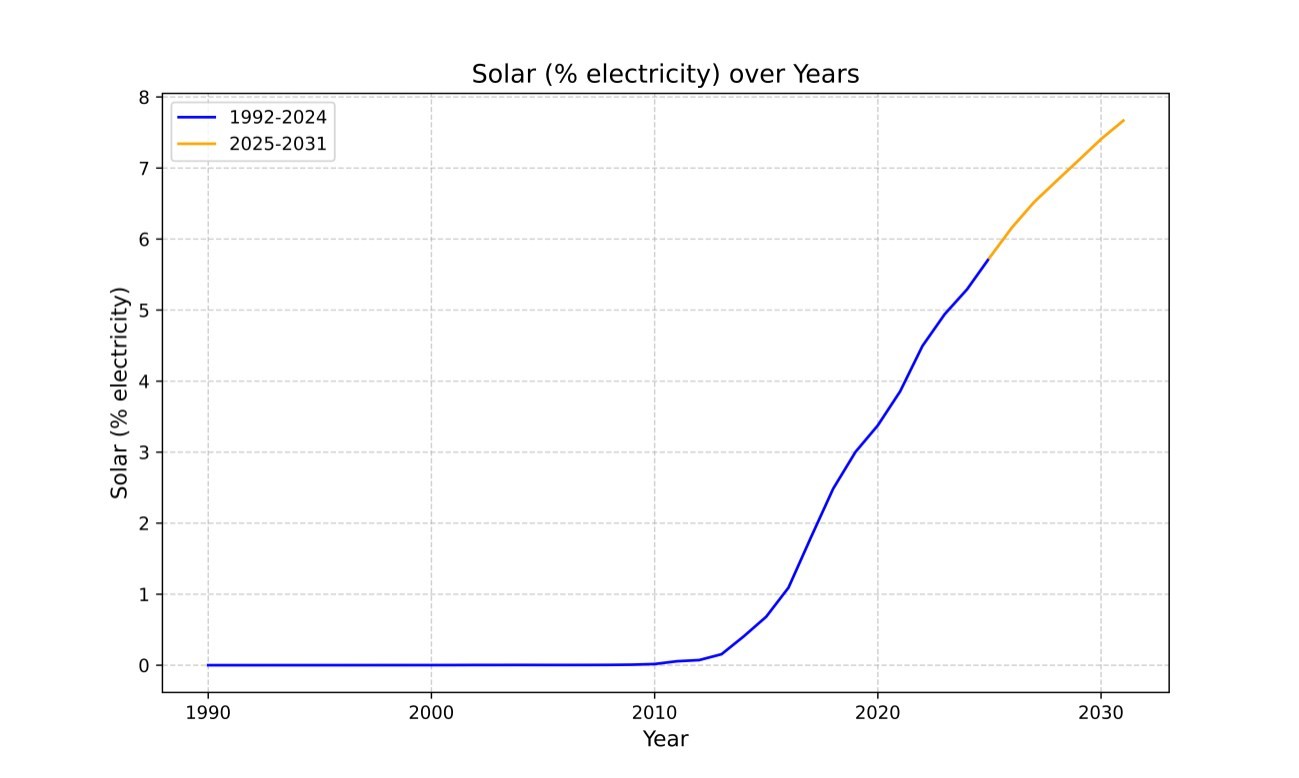}
\caption{Solar (\% Electricity) Prediction Scenario.}
\label{fig12}
\end{figure}

\textit{Solar (\% Electricity):} The energy from solar shows an exponential behavior where there are few contributions below 2010 and substantial increase above 2020 in Fig. \ref{fig12}. It forecasts further exponential growth to beyond 7 percent in 2031. The fact, this trend represents expresses the fast-growing application of solar technology and the decreasing price.

\begin{figure}[!t]
\centering
\includegraphics[width=0.48\textwidth]{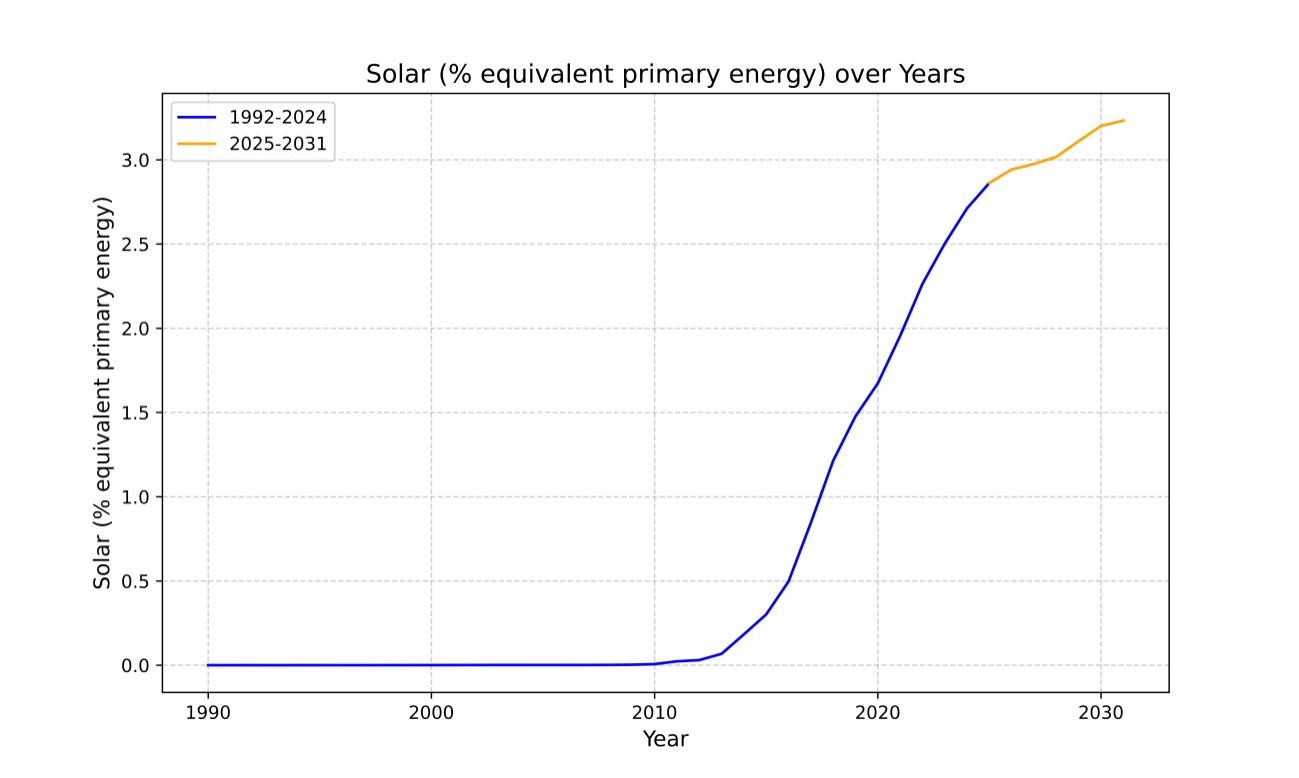}
\caption{Solar Energy (\% equivalent primary energy).}
\label{fig13}
\end{figure}

\textit{Solar Energy (\% equivalent primary energy):} Fig. \ref{fig13} correlates with the electricity graph where contributions are insignificant prior to 2010 but steep thereafter, and insignificant post-2020. Solar is projected to exceed 3\% in equivalent primary energy share by 2031 and is critical to the process of energy transition.

\subsection{SARIMA Prediction Scenario}

\begin{figure}[!t]
\centering
\includegraphics[width=0.48\textwidth]{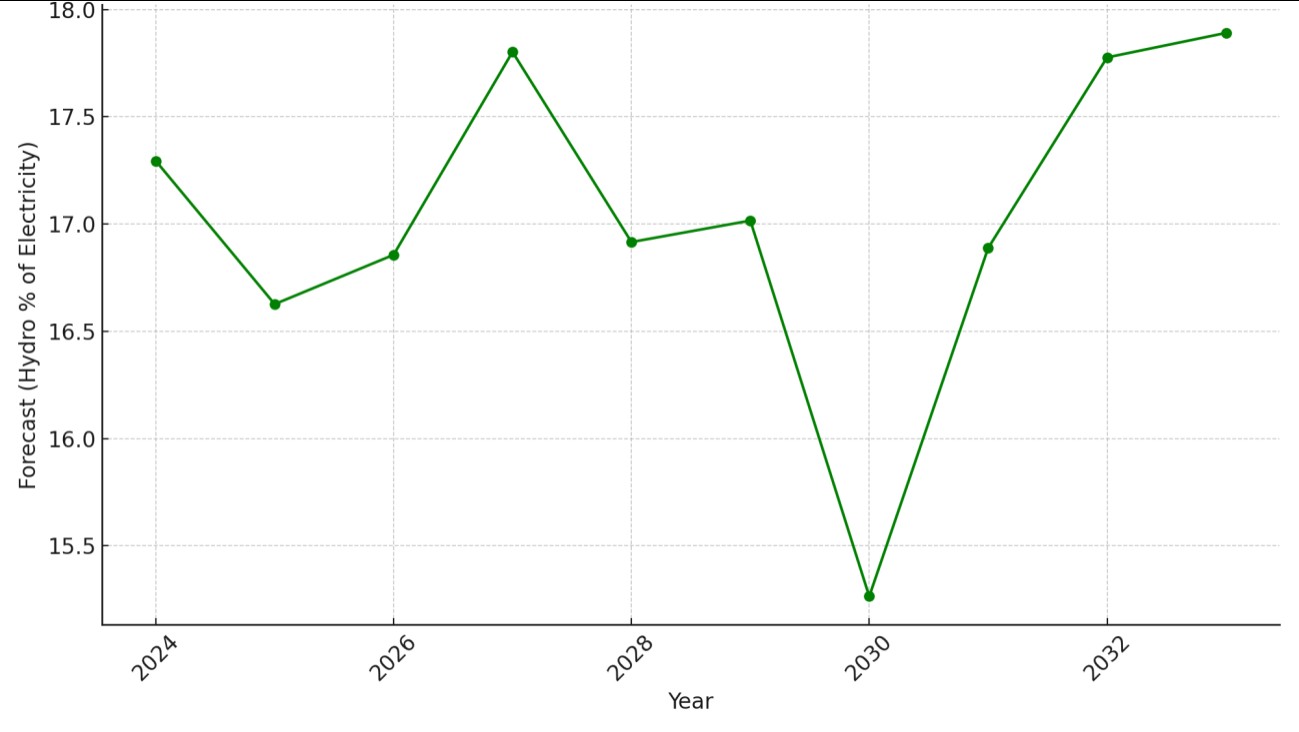}
\caption{SARIMA based Forecast from Hydro (\% of electricity).}
\label{fig14}
\end{figure}

\textit{Hydro electricity usage forecast:} Fig. \ref{fig14} gives the projection on the percentage contribution of hydroelectricity to the total electricity consumption between the years 2024 and 2032. From the SARIMA analysis, the contribution of hydroelectricity is expected to reduce from 17.0\% in 2024 to 15.5\% in 2030 before increasing to 18.0\% in 2032. These sources also illustrate the current, periodic and erratic nature of hydroelectricity due to natural factors like availability of water and climatic conditions. The increase from 2030 indicates the considerations that policy interventions or infrastructure additions could be used to stabilize and boost hydroelectric power production. The model represents these fluctuations adequately in relation to the variability of hydroelectricity in the overall portfolio of renewable energy sources.

\begin{figure}[!t]
\centering
\includegraphics[width=0.48\textwidth]{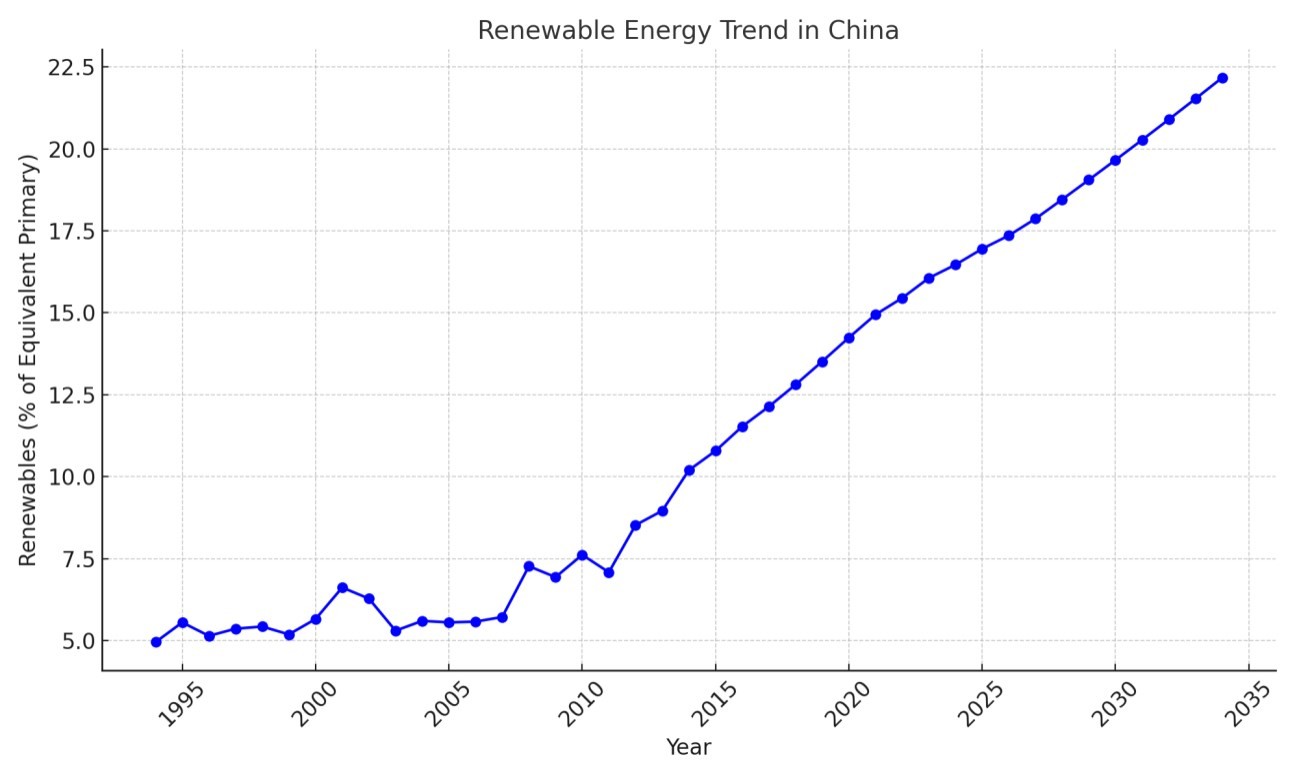}
\caption{Renewable energy Trend in China.}
\label{fig15}
\end{figure}

\textit{Renewable energy Trend in China:} Fig. \ref{fig15} shows the change in renewable energy as a percent of EPE in China in the range 1995 to 2035. Analysed historical data shows that the contribution of renewable energy sources has been rather constant and low before 2010 and then, the rate has constantly been rising sharply from 2010 onwards. The SARIMA model has predicted this growth to continue and therein reaching 22.5\% of equivalent primary energy per renewable in 2035. This exponential trend suggests the efficiency of this aggressive approach of China to invest in renewal energy systems and technology which includes large investments on silicon solar, wind, hydro power facilities. The observed increase indicates the country's overall positive dynamics of moving to a low-carbon energy system and achieving the objectives of decarbonization.

\vspace{0.01cm}
\begin{table}[!t]
\caption{Error Metrics Analysis Output of Different Parameters (KNN)}
\label{tab1}
\centering
\small
\begin{tabular}{lcc}
\toprule
\textbf{Parameter} & \textbf{MSE} & \textbf{MAE} \\
\midrule
Renewables (\% equiv. primary) & 0.313 & 0.447 \\
Renewables (\% electricity) & 0.914 & 0.770 \\
Hydro (\% equiv. primary energy) & 0.333 & 0.491 \\
& & \\
Wind (\% equiv. primary energy) & 1.896 & 0.664 \\
Wind (\% electricity) & 4.317 & 1.051 \\
Solar (\% equiv. primary energy) & 0.087 & 0.195 \\
Solar (\% electricity) & 0.484 & 0.436 \\
\bottomrule
\end{tabular}
\end{table}

\subsection{Error Metrics Analysis}
The error metrics table evaluates KNN model forecasts for solar, wind, and hydro. Solar energy projections are most accurate with MSE = 0. Hydro energy estimates are less accurate, particularly for electricity contributions. Errors in wind energy projections are lower for equal primary energy but higher for intermittent electricity. The KNN model covers solar energy development, but lacks sufficient precision for wind and hydro power dynamics.

The SARIMA model accurately predicts renewable energy trends with low mistakes. In renewables, MAE (0.57), MSE (1.13), and RMSE (1.06) indicate a high match, whereas MAPE (9.11\%) confirms an average deviation of less than 10\%. Results show continual increase of renewables due to policy and technology developments. Hydroelectricity estimates indicate a somewhat higher MAE (1.58), moderate MSE (1.01), and reasonable MAPE (6.64\%). While SARIMA lacks RMSE for hydro, it effectively captures seasonal variations. SARIMA provides accurate long-term renewable energy forecasts and aids policy planning.

\begin{table}[!t]
\caption{Error Matrices Analysis for SARIMA Prediction}
\label{tab2}
\centering
\small
\begin{tabular}{lcccc}
\toprule
\textbf{Parameter} & \textbf{MAE} & \textbf{MSE} & \textbf{RMSE} & \textbf{MAPE} \\
\midrule
Renewables (\% equiv. primary) & 0.57 & 1.13 & 1.06 & 9.11 \\
Hydro (\% equiv. primary) & 0.29 & 0.42 & -- & 5.42 \\
\bottomrule
\end{tabular}
\end{table}

\section{Conclusion}
This research employs KNN and SARIMA, state-of-the-art machine learning models, to anticipate Chinese renewable energy generation trends.  Through a large dataset of over three decades, the suggested models show high predictive skills that could help renewable energy systems overcome intermittency, non-linearity, and seasonality.  SARIMA forecasts seasonal and temporal trends well, while KNN makes good localised predictions using neighbouring data sets.  It shows substantial solar and wind energy growth trends that match China's ambitious renewable energy policies and technological advances.  Due to resource constraints, hydroelectricity is expected to stabilise, while renewable energy policy is shifting.  Error metric study shows model reliability and predictive accuracy with low MSE and MAE values.  This has major consequences for policy planners, grid managers, and investors for strategic decisions, resource planning, and infrastructure development.  Data-driven forecasting frameworks can boost renewable energy integration into the national grid while ensuring energy management sustainability and efficiency.  Adding a hybrid model and considering forthcoming renewable technologies can increase prediction accuracy.

\end{document}